\documentclass[11pt]{article}

\usepackage[english]{babel}

\usepackage{authblk}
\usepackage{amsfonts}
\usepackage{amssymb}
\usepackage{amsthm}
\usepackage[braket, qm]{qcircuit}
\usepackage{braket}
\usepackage{verbatim}
\usepackage{multicol}
\usepackage{rotating}
\usepackage{hhline}
\usepackage[table,x11names]{xcolor} 
\definecolor{vlightgray}{cmyk}{0,0,0,0.125}

\usepackage{tikz}
\usetikzlibrary{arrows.meta}

\usepackage[letterpaper,top=2cm,bottom=2cm,left=3cm,right=3cm,marginparwidth=1.75cm]{geometry}

\usepackage{amsmath}
\usepackage{graphicx}
\usepackage[colorlinks=true, allcolors=blue]{hyperref}

\title{Universal Statistical Simulator}
\author[1]{Mark Carney*}
\author[2]{Ben Varcoe}
\affil[1]{Quantum Threat Group, Cybersecurity Research, Santander Global, London}
\affil[ ]{\href{mailto:mark.carney@gruposantander.com}{mark.carney@gruposantander.com}}
\affil[2]{Department of Physics and Astronomy, University of Leeds, Leeds, United Kingdom}
\affil[ ]{\href{mailto:b.varcoe@leeds.ac.uk}{b.varcoe@leeds.ac.uk}}

\begin{document}

\maketitle

\begin{abstract}
The Quantum Fourier Transform is a famous example in quantum computing for being the first demonstration of a useful algorithm in which a quantum computer is exponentially faster than a classical computer. 
However when giving an explanation of the speed up, understanding computational complexity of a classical calculation has to be taken on faith. 
Moreover, the explanation also comes with the caveat that the current classical calculations might be improved. 
In this paper we present a quantum computer code for a Galton Board Simulator that is exponentially faster than a classical calculation using an example that can be intuitively understood without requiring an understanding of computational complexity. 
We demonstrate a straight forward implementation on a quantum computer, using only three types of quantum gate, which calculates $2^n$ trajectories using $\mathcal{O} (n^2)$ resources. 
The circuit presented here also benefits from having a lower depth than previous Quantum Galton Boards, and in addition, we show that it can be extended to a universal statistical simulator which is achieved by removing pegs and altering the left-right ratio for each peg.

\end {abstract}

\section{Introduction}
The quantum mechanical Galton Board has now been a core concept in quantum computing for nearly a decade \cite{Aaronson2014} with its roots extending back to the development of linear optics quantum computing.
Apart from the interest in demonstrating quantum enhanced performance, a Galton Board (GB) is also interesting for its computational properties. Removing the pegs and altering the balance of each peg allows for statistical calculations to be performed. In this way, the GB is a statistical simulator rather than a universal computer, nevertheless it allows for a number of applications\cite{Kadian2021} from complex system simulation\cite{Bouwmeester1999,Bottcher2021,Max2022,Bagrets2021}, modelling random walks across graphs\cite{Nguyen2021,Roehsner2021,Lahini2018,Su2019,Blank2021}, machine learning\cite{Li-HuaLu,Dernbach2019,Souza2019}, modelling stock price fluctuations\cite{Fama1965,emmanoulopoulos2022quantum,yi2022information,pistoia2021quantum}, one way functions for cryptography\cite{AbdEl-Latif2021,El-Latif2021,Janani2021,El-Latif2020}, and sampling and search problems\cite{Aaronson2014,Avrachenkov2018,Xia2020,Shenvi2003,Segawa2021}.

In this paper we present an intuitive approach to generating a Quantum Galton Board (QGB) that mimics the physical action of the classical GB in a quantum circuit. 
The aim of this paper is not necessarily to present a new quantum algorithm, but rather to provide a clear and transparent demonstration of the exponential speed up that can be provided with a quantum processor.
In comparison with previous research which replicated the output of the classical GB, we create a quantum circuit that creates a superposition of all possible trajectories. 
In a "sampling" mode this possesses the statistics of a GB, but it also allows for individual trajectories to be tailored to meet specific requirements. 
Moreover, in contrast with previous work the entire circuit is performed using combinations of three Clifford operations and ancilla qubits. To perform an n-level QGB we require $n(2n-1)$ gates and $2n$ qubits ($2n-1$ working qubits, and a recycled control qubit). 

The requirement of a large number of ancilla qubits poses a large overhead. however, as the aim is to presenting an approach that generates the actual trajectories in a classical GB this is a relatively small price to pay.

\section{Preliminaries}

\subsection{The classical Galton Board}

\begin{figure}
    \centering
    \includegraphics{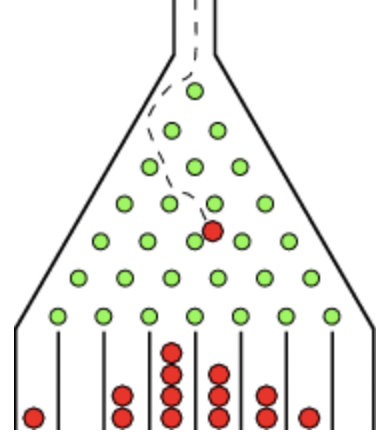}
    \caption{A standard pyramidal Galton board, from Wolfram MathWorld \cite{weisstein}.}
    \label{fig:gbwolf}
\end{figure}

The standard construction of a Galton board is presented in figure \ref{fig:gbwolf}. 
In the regular formulation, each column from $1 \ldots n+1$ of the output of an $n$ level Galton board has, for $q = 1-p = 0.5$ the value 
\begin{align}\label{eqn:count}
    \binom{n}{k} p^k q^{n-k} = \frac{1}{2^n} \binom{n}{k} 
\end{align}
Although it is omitted here, the proof that a Galton board produces the normal distribution can be obtained via the De Moivre-Laplace theorem.

\section{The Quantum Galton Board}

The basis for our approach to the construction of the QGB is to model the action of an individual peg on a physical Galton board. In the classical GB an ideal ball hitting the ideal peg produces a 50\% probability of going left or right, then hitting another peg on the board.

To mimic this, we take a set of $\ket{0}$ initialized qubits, and invert the middlemost qubit using an $X$ gate. We then construct a series of superposed $SWAP$s that this `ball' can effectively `fall through' mimicking the action of a real ball through an array of pegs on a Galton board.

We can use this model to get a normal distribution by relying on equation \ref{eqn:count} in the same way that you do on a physical, macro-scale Galton board.

\subsection{A Quantum Peg}

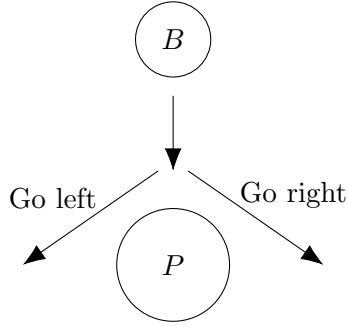
\begin{figure}
    \centering
\begin{tikzpicture}
    \draw (0,0) circle (0.5) node {$B$};
    \draw (0,-3) circle (0.75) node {$P$};
    \draw [-{Latex[length=3mm]}] (0,-0.75) -- (0,-1.75);
    \draw [-{Latex[length=3mm]}] (0.2,-1.75) -- (2,-3) node[above, midway, xshift=0.5cm] {Go right};
    \draw [-{Latex[length=3mm]}] (-0.2,-1.75) -- (-2,-3) node[above, midway, xshift=-0.5cm] {Go left};
\end{tikzpicture}
    \caption{Basic model of a ball $B$ interacting with a physical peg $P$ that forms the basis of our `quantum peg' circuit module in fig \ref{fig:quantum_peg}.}
    \label{fig:classical_peg}
\end{figure}

\begin{figure}
    \centering
    \[
    \Qcircuit  {
    & q_0 & \lstick{\ket{0}} & \gate{H}  & \ctrl{1}      &  \targ     & \ctrl{2}     & \qw &\qw\\
    & q_1 & \lstick{\ket{0}} & \qw       & \qswap        &  \qw       &  \qw         & \meter&\cw \\
    & q_2 & \lstick{\ket{0}} & \gate{X}  & \qswap \qwx   &  \ctrl{-2} &  \qswap      & \qw &\qw\\
    & q_3 & \lstick{\ket{0}} & \qw       & \qw           & \qw        &  \qswap \qwx & \meter&\cw 
    } \]
    \caption{A Quantum circuit analogue of the physical peg in figure \ref{fig:classical_peg}. Qubit $q_2$ is the input channel or `ball' and meters (measurements) indicate the output channels of the `quantum peg'. }
    \label{fig:quantum_peg}
\end{figure}
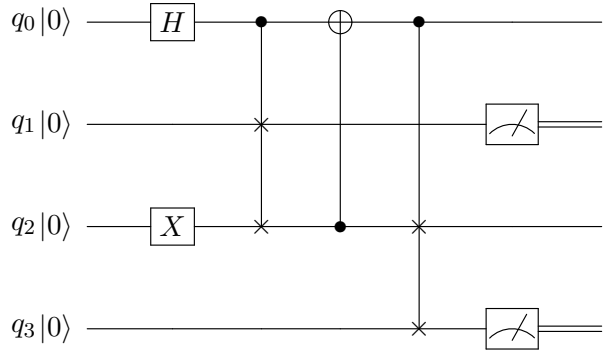

We first construct a `quantum peg' module (figure \ref{fig:quantum_peg}) that we can then replicate to build a complete Galton board.
The basic peg uses three working qubits ($q_1$, $q_2$, and $q_3$) and one control qubit ($q_0$). 
We initialize all qubits to zero, then place the control qubit $q_0$ into superposition with the Hadamard gate and place a `ball' on the middle working qubit, $q_2$ using an $X$-gate.

We then perform a controlled-$SWAP$ operation on $q_1$ and $q_2$ giving us an eventual state for $\ket{q_2 q_1 q_0}$ as $$ \frac{1}{\sqrt{2}} ( \ket{011} + \ket{100} ) $$ which by an application of an inverted $CNOT$ on $q_2$ to $q_0$ become $$ \frac{1}{\sqrt{2}} ( \ket{011} + \ket{101} ) $$ Given our control qubit is now a stable $\ket{1}$ we can apply a swap between $q_2$ and $q_3$ to give the desired final state of $$ \ket{q_3 q_2 q_1 q_0} =  \frac{1}{\sqrt{2}} ( \ket{0011} + \ket{1001} ) $$ 

This module can, with some care around the control qubit, be replicated to reproduce each level of the Galton board successively. 

\subsection{A 3-peg QGB}

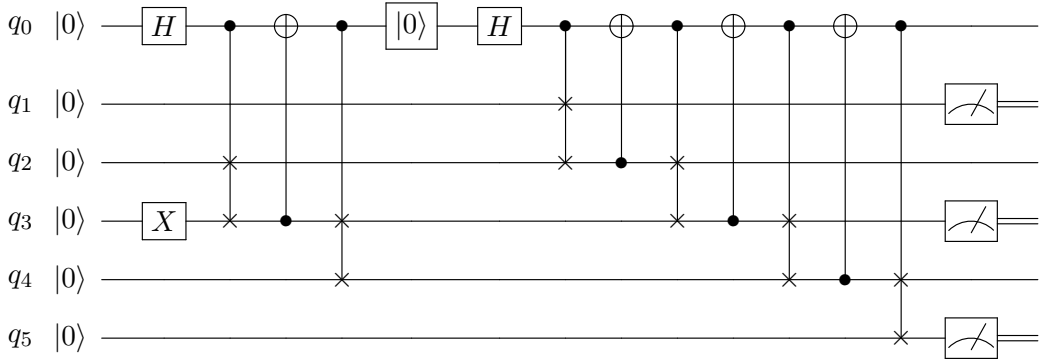
\begin{figure}
    \centering
    \[
    \Qcircuit @C=1.4em @R=1.2em {
    & q_0 & 
        & \lstick{\ket{0}} & \gate{H}  & \ctrl{2}      &  \targ     & \ctrl{3}     & \gate{\ket{0}} 
        & \gate{H}  
        & \ctrl{1}    & \targ     & \ctrl{2}    & \targ     & \ctrl{3}      
        & \targ     & \ctrl{4}    & \qw     & \qw \\
    & q_1 & 
        & \lstick{\ket{0}} & \qw       & \qw           & \qw        & \qw          & \qw 
        &  \qw      
        & \qswap \qwx & \qw       & \qw         & \qw       & \qw           
        & \qw       & \qw         & \meter  & \cw \\
    & q_2 & 
        & \lstick{\ket{0}} & \qw       & \qswap        &  \qw       &  \qw         & \qw 
        &  \qw      
        & \qswap \qwx & \ctrl{-2} & \qswap \qwx & \qw       & \qw           
        & \qw       & \qw         & \qw     & \qw \\
    & q_3 & 
        & \lstick{\ket{0}} & \gate{X}  & \qswap \qwx   &  \ctrl{-3} &  \qswap      & \qw 
        &  \qw      
        & \qw         & \qw       & \qswap \qwx & \ctrl{-3} & \qswap    
        & \qw       & \qw         & \meter  & \cw \\
    & q_4 & 
        & \lstick{\ket{0}} & \qw       & \qw           & \qw        &  \qswap \qwx & \qw 
        & \qw       
        & \qw         & \qw       & \qw         & \qw       & \qswap \qwx   
        & \ctrl{-4} & \qswap      & \qw     & \qw \\
    & q_5 & 
        & \lstick{\ket{0}} & \qw       & \qw           & \qw        & \qw          & \qw
        & \qw      
        & \qw         & \qw       & \qw         & \qw       & \qw 
        & \qw       & \qswap \qwx & \meter  & \cw 
} \]
    \caption{A Quantum circuit analogue of the physical peg in figure \ref{fig:classical_peg}.}
    \label{fig:quantum_3peg}
\end{figure}

We detail in figure \ref{fig:quantum_3peg} a 3-peg (2 level) version of a QGB implemented using our method. Note, for 6 qubits, we get 3 classical bits out, which is half what is achieved by other methods (see \cite{Rattew2021}). However, we have what we consider to be close to a minimal circuit depth for our algorithm. 

The circuit in fig \ref{fig:quantum_3peg} can be thought of in three sections. The first is identical to the individual peg in fig \ref{fig:quantum_peg}. We then reset the control qubit $q_0$ and proceed to take the `output' of the first peg on qubits $q_2$ and $q_4$, and apply the same quantum peg logic, adding an additional inverted $CNOT$ (the 10th gate column in our diagram fig \ref{fig:quantum_3peg}). This $CNOT$ now returns the control qubit probability to 50\%, which then allows us to proceed onto implementing a quantum peg for the other `half' of the circuit. Finally, we have three measurements on $q_1$, $q_3$, and $q_5$, with state of measurement given by $$ \ket{q_5 q_3 q_1} = \frac{1}{\sqrt{4}} (\ket{001} + 2 \ket{010} + \ket{100}  ) $$ with the full state begin given by $$ \ket{q_5 q_4 q_3 q_2 q_1 q_0 } = \frac{1}{2} ( \ket{10000} + 2 \ket{00100} + \ket{00001} ) \sqrt{3}\ket{1} $$

\subsection{Scaling the QGB}

The additional $CNOT$ after our peg module circuit we described previously is needed to re-balance the control qubit for the successive gates. The requirement for a mid-circuit reset on the control qubit is also required for every successive level of the circuit. Thus, we can expect for every `peg' we want to model, we require at most 4 gates, with optional local optimizations of only needing 3-gates. 

\subsection{QGB Features}\label{sec:features}

We first note that our outputs will differ from other approaches, as each output measurement will have precisely a single `1' output. Thus, some post-processing should be required to determine the final output.

We require $n$-many ancilla qubits for each $n$ bits of desired output, which is a limitation given the current scale of NISQ devices that are currently available. 

In terms of gate count, the following give the upper bound for $n$ desired output bits, subject of course to local optimisations:

\begin{itemize}
    \item We require $n$ Hadamard gates, and $n$ reset ($\ket{0}$) gates for a Galton board of $n$-many levels, plus one $X$ gate for the `ball'.
    \item We require 4 gates per `peg', the total number of which is governed by triangle numbers.
    \item We require $n+1$ measurement gates. 
\end{itemize}

This gives us an upper limit on our gates of $$ n+n+ 4 \Big( \frac{n^2+n}{2} \Big) + n + 1 + 1 = 2n^2+5n+2 $$ We note that for $n+1$-many output bits, this equates to $n$ levels of a Galton board. 

Our method has circuit depth less than half that demonstrated in \cite{Rattew2021}, for example, who present 167 gates on 5-qubit output, compared to our upper limit of 76. However, ours requires post-processing to convert it from an array of $n$ bits with a single 1 to a full 5-bit normally distributed output. The significant decrease on circuit depth, however, significantly reduces the amount of error our circuits are prone to experiencing, meaning in part that an increase in the number of runs does not necessitate an increase in the level of error.

\section{Experimental Results for QGB}

We performed local and remote simulations using Qiskit, IBM's python-based quantum circuit library.

\subsection{Remote Simulations}

\begin{figure}
    \centering
    \includegraphics[width=0.6\textwidth,keepaspectratio]{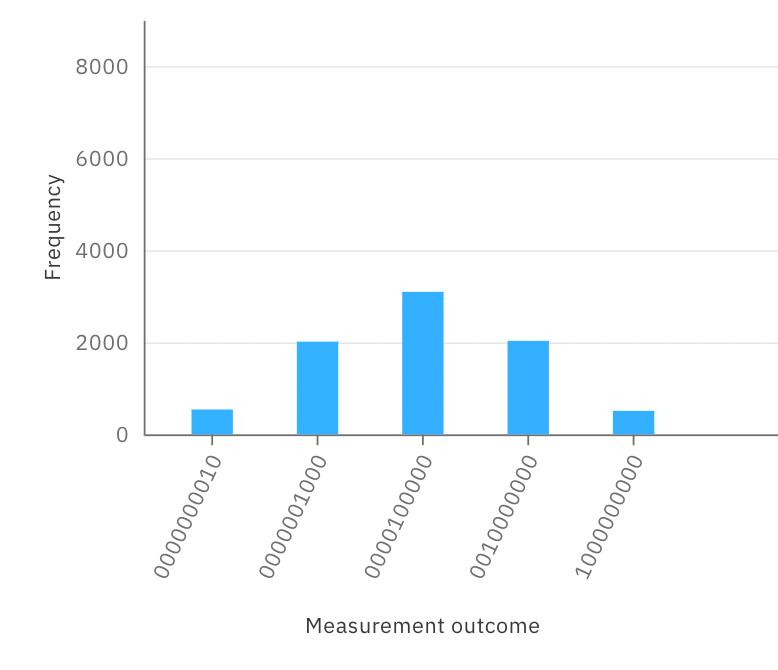}
    \caption{Output results from an IBM simulation of the 4-QGB circuit.}
    \label{fig:ibm_histogram}
\end{figure}

We produced a 4-level Quantum Galton Board (4-QGB) circuit and implemented this on the IBM-QX quantum simulator, as access to quantum computers with more than 5 real qubits is currently restricted. The IBM-QX simulator, however, uses quantum random numbers for its simulations, so the fidelity to actual experiment is likely as close as can be achieved.

We constructed the circuit in OpenQASM - listed in Appendix \ref{appA}. The IBM transpiler made no alterations to our code, and simulated the circuit, see figure \ref{fig:ibm_4QGB} in Appendix \ref{appB}. 

The results are listed in the histogram in figure \ref{fig:ibm_histogram}. As you can see, the expected normal distribution can be found, albeit on 5 output qubits. 

\subsection{Real Simulations of Quantum Peg}

\begin{figure}
    \centering
    \includegraphics[width=0.6\textwidth,keepaspectratio]{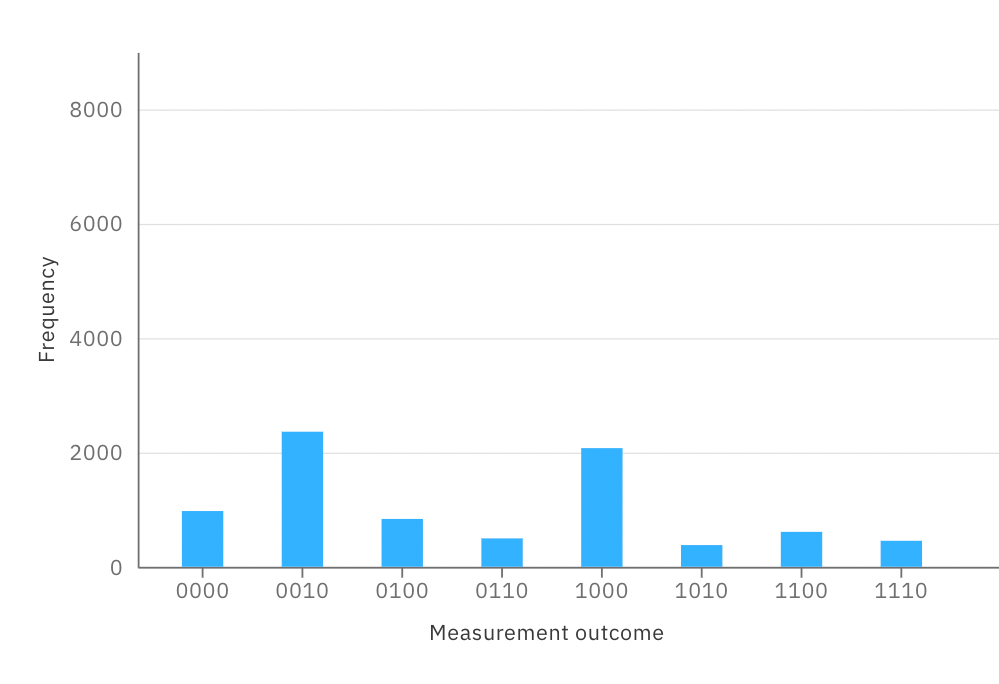}
    \caption{Results histogram from a quantum computer hardware run of the quantum peg circuit (fig \ref{fig:quantum_peg})}
    \label{fig:ibm_peg_results}
\end{figure}

\begin{table}[b]
\centering
\begin{tabular}{|c|c|c|}
\hline
State       & Num. & \% (from 8192 shots) \\ \hhline{|=|=|=|}
$\ket{000}$ & 975 & $\approx 11.90$     \\ \hline
\rowcolor{vlightgray} $\ket{001}$ & 2362 & $\approx 28.83$     \\ \hline
$\ket{010}$ & 837  & $\approx 10.22$     \\ \hline
$\ket{011}$ & 497  & $\approx 6.07$      \\ \hline
\rowcolor{vlightgray} $\ket{100}$ & 2075 & $\approx 25.33$     \\ \hline
$\ket{101}$ & 380  & $\approx 4.64$      \\ \hline
$\ket{110}$ & 611  & $\approx 7.46$      \\ \hline
$\ket{111}$ & 455  & $\approx 5.55$      \\ \hline
\end{tabular}
\caption{Results from running the QGB peg circuit on quantum computer hardware. The desired outcomes are highlighted in grey.\label{tab:qgb-results}}
\end{table}

Although simulating a full QGB is not immediately possible, it was possible to simulate a quantum peg. The circuit was uploaded to IBM-QX and the transpiler created the circuit depicted in figure \ref{fig:ibm_transpiled_peg}. 

This transpiled circuit, however, clearly has a lot of noise owing to the fact that 5 active quantum gates have now become 64 gates, which necessarily leads to a significant increase in error. The results are presented in figure \ref{fig:ibm_peg_results} and table \ref{tab:qgb-results}.

As is apparent, the output is very noisy, as the desired results represent only 54.16\% of the total output, with the rest being noise. That said, the two peaks are on the desired outputs we defined earlier, $\ket{100}$ and $\ket{001}$.\footnote{\emph{NB} - the measurement operators were not applied to qubit $q_0$ during the runs, and IBM-QX defaults to a `0' readout in that case.}

\subsection{Local Simulation and Rescaling of QGB Output}\label{sec:qgb-locsim}

\begin{figure}
    \centering
    \includegraphics[width=0.8\textwidth]{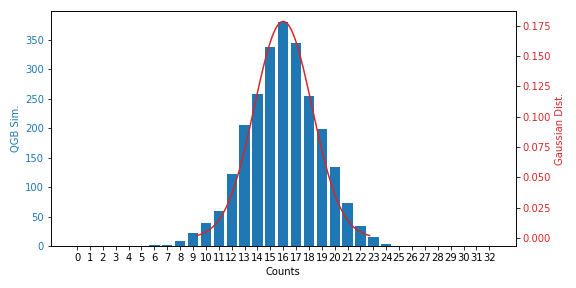}
    \caption{The re-scaled output of the 4-QGB locally simulated circuit. The plot here shows the output of our processing on the simulated QC output against the normal distribution with parameters $var=5$ (for visibility) and $\mu = 16$.}
    \label{fig:qgb_local_hist}
\end{figure}

To demonstrate the post-processing we mentioned in section \ref{sec:features}, we loaded the circuit onto a local simulator and generated 20,000 shots, storing each result. For the outputs we reassigned them with an integer from 0 to 4 (`00001' became `0', `00010' became `1', and so on). Under this substitution, the mean average was 1.9977, the standard deviation was 1.001521, and the variance was 1.002995.

With this reassignment done, we took blocks of 8 outputs and summed each block. This scaled our output from 0 to 32, as the maximum sum is $8\times 4$. We then plotted how many of each sum we had, and compared this to the normal distribution. The result can be seen in figure \ref{fig:qgb_local_hist}.

As can be seen, we have successfully scaled our output of individual bits under normal distribution to a range - this method can be repeated for any multiple of the highest-value substitution. 

\section{A Biased QGB}

Of course, regular normal distributions are only part of the story when it comes to generating useful random numbers. We present a construction of a `biased QGB` (B-QGB) where we can execute control over the final distirbution with significant accuracy. 

\subsection{The Biased Quantum Peg}

\begin{figure}
    \centering
    \[
    \Qcircuit  {
    & q_0 \text{ } & \lstick{\ket{\psi}} & \gate{\ket{0}} & \gate{R_x(\theta)}  & \ctrl{1}      &  \targ     & \ctrl{2}     & \qw \\
    & q_1 & \lstick{\ket{0}} & \qw       & \qw & \qswap        &  \qw       &  \qw         & \qw \\
    & q_2 & \lstick{\ket{0}} & \qw & \gate{X}  & \qswap \qwx   &  \ctrl{-2} &  \qswap      & \qw \\
    & q_3 & \lstick{\ket{0}} & \qw & \qw       & \qw           & \qw        &  \qswap \qwx & \qw 
} \]
    \caption{A biased quantum peg, based on figure \ref{fig:classical_peg} with the $H$ gate replaced with an $x$-axis rotation $R_x(\theta)$}
    \label{fig:biased_peg}
\end{figure}
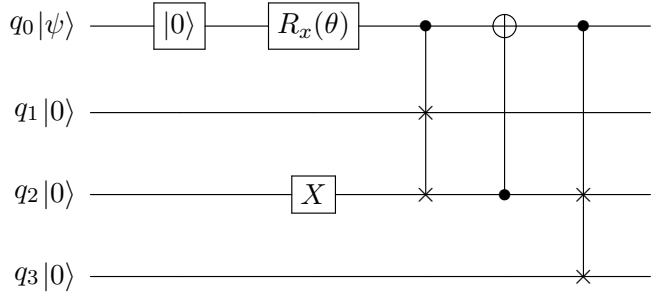

As before, we follow a modular approach and construct a quantum peg circuit that is biased, depicted in figure \ref{fig:biased_peg}.

This circuit follows naturally from the observation that our circuit in figure \ref{fig:quantum_peg} can have its $H$ gate replaced with a rotation about the $x$-axis, a gate usually denoted $R_x(\theta)$ which is defined as \begin{equation*} R_x(\theta) = 
    \begin{pmatrix}
        \cos{\frac{\theta}{2}} & -i \sin{\frac{\theta}{2}} \\
        -i \sin{\frac{\theta}{2}} & \cos{\frac{\theta}{2}}
    \end{pmatrix}
\end{equation*}
As such, if we replace the $H$ gate with $R_x(\frac{\pi}{2})$, then the resultant circuit operates precisely the same - the $\cos{\frac{\pi}{4}}$ terms replace the $(\frac{1}{\sqrt{2}} - \frac{1}{\sqrt{2}})$ null factors on the diagonal. Likewise, the $- \frac{i}{\sqrt{2}}$ and $\frac{1}{\sqrt{2}}$ terms are equivalent under the regular norm used in measurement.

With this, we can then see that by introducing a free rotation about the $x$-axis in our circuit we can gain more control over the nature of the underlying probabilities affecting the `quantum peg'. 

However, to manage possibly unique rotations per peg, we also require an additional prefix of a $RESET$ gate, which is theoretically no problem but might present some issues on certain quantum computers. However, the payoff of distribution control, and the fact that our circuits are significantly shorter still, means that this payoff is worthwhile. 

The overall equation for this biased quantum peg is then given as follows. First note that: $$ R_x(\theta) \ket{0} = 
\begin{pmatrix}
    \cos{\frac{\theta}{2}} \\
    - i \sin{\frac{\theta}{2}}
\end{pmatrix}$$ and the complimentary rotation $$ 
\overline{R_x}(\theta) \ket{0} = R_x(\pi - \theta) \ket{0} =
\begin{pmatrix}
    \cos{\frac{\pi - \theta}{2}} \\
    - i \sin{\frac{\pi - \theta}{2}}
\end{pmatrix}
$$
We shall denote these states as $\ket{R_\theta}$ and $\ket{\overline{R_\theta}}$ respectively. Thus the state of the biased quantum peg in figure \ref{fig:biased_peg} is given by 
\begin{equation}\label{eqn:bqp-state}
\ket{q_3 q_2 q_1 q_0} = ( \ket{001 R_\theta} + \ket{100 \overline{R_\theta}} ) 
\end{equation}

The way that these states feed forwards into successive rows on our QGB means that we can have a given rotational bias on each row that will behave as we expect under successive compositions.

\subsection{Biased Quantum Peg Example}\label{sec:bqpeg-ex}

We provide the following worked example for a biased peg that has the probability of measuring a 1 on the upper qubit of 75\% and 25\% on the lower qubit - qubits $q_1$ and $q_3$ respectively in figure \ref{fig:biased_peg}. 

To do this we set our $\theta = \frac{2 \pi}{3}$ and note that the final state as given in equation \ref{eqn:bqp-state} becomes 
\begin{equation*}
\ket{q_3 q_2 q_1 q_0} = \Big( \sqrt{\frac{3}{4}}\ket{0011} + \frac{1}{\sqrt{4}}\ket{1001} \Big) 
\end{equation*}
When we apply measurement, these terms will give us the desired output of 75\% chance of a $\ket{0011}$ state, and a 25\% chance of a $\ket{1001}$ state. Should we wish/require to, we can flip this order by letting $\theta = \frac{\pi}{3}$, as we might expect.

\subsection{Biased QGB Features}

The overall gate count does not increase too significantly. If we suppose that we want per peg row bias to be set, which is the maximal case for control of the circuit in our current setup, then we have:
\begin{itemize}
    \item 5 gates per peg - 
    \begin{itemize}
        \item one $RESET$, 
        \item one $R_x(\theta)$, 
        \item two controlled-$SWAP$s,
        \item (up to) two $CNOT$.
    \end{itemize}
    \item $n+1$ measurement gates.
    \item One $X$ gate for the `ball'.
\end{itemize}

Again, we use the triangle numbers to get a maximum gate count for an $n$-level QGB given by:
\begin{equation}
    3 ( n^2 + n) + n + 2
\end{equation}

In section \ref{fig:quantum_3peg_fine_biased} we take this idea further and demonstrate fine-grained control of each individual peg without adding extra ancilla qubits.

\section{Experimental Results for Biased-QGB}

\subsection{Biased-QGB Peg on QC Hardware}

\begin{table}
\centering
\begin{tabular}{|c|c|c|}
\hline
State       & Num. & \% (from 8192 shots) \\ \hhline{|=|=|=|}
$\ket{000}$ & 1108 & $\approx 13.52$     \\ \hline
\rowcolor{vlightgray} $\ket{001}$ & 2991 & $\approx 36.51$     \\ \hline
$\ket{010}$ & 851  & $\approx 10.39$     \\ \hline
$\ket{011}$ & 490  & $\approx 5.98$      \\ \hline
\rowcolor{vlightgray} $\ket{100}$ & 1464 & $\approx 17.87$     \\ \hline
$\ket{101}$ & 492  & $\approx 6.00$      \\ \hline
$\ket{110}$ & 352  & $\approx 4.30$      \\ \hline
$\ket{111}$ & 444  & $\approx 5.42$      \\ \hline
\end{tabular}
\caption{Results from running a biased QGB peg on quantum computer hardware. The desired outputs are highlighted in grey.\label{tab:bqgb-results}}
\end{table}

\begin{figure}
    \centering
    \includegraphics[width=0.6\textwidth,keepaspectratio]{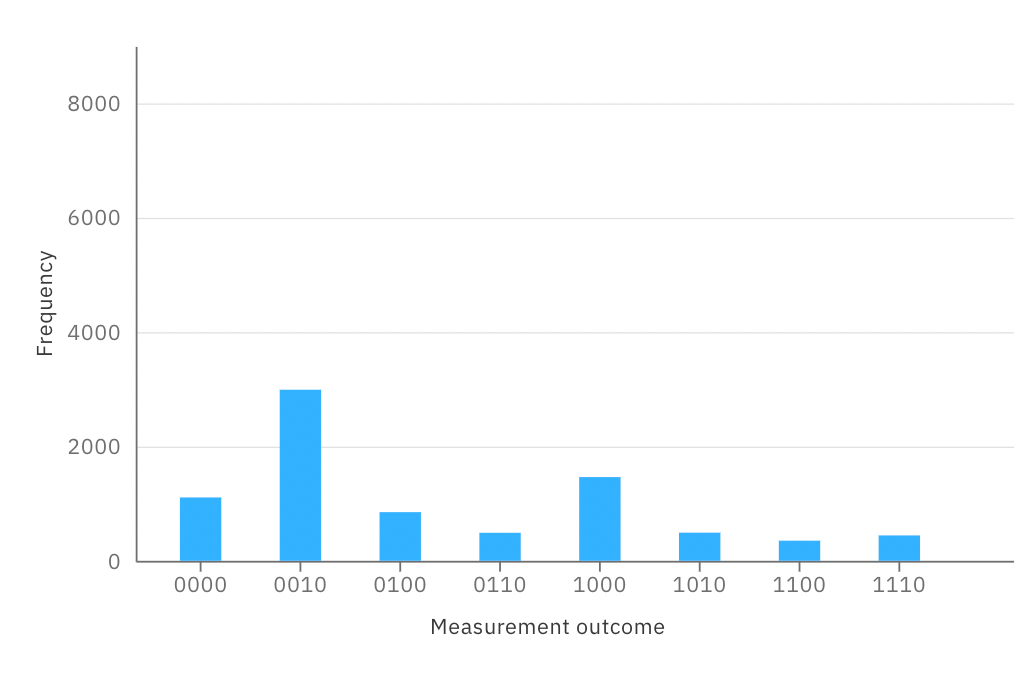}
    \caption{Results histogram from a real QC run of the biased quantum peg circuit (fig \ref{fig:biased_peg})}
    \label{fig:ibm_biased_results}
\end{figure}

We ran the biased quantum peg circuit from fig \ref{fig:biased_peg} on the \verb|ibmq-manila| quantum computer. The transpiled circuit was of similar depth to the one generated previously, also shown in figure \ref{fig:ibm_transpiled_peg} - as we can see, it is broadly the same. The results are presented in the graph in figure \ref{fig:ibm_biased_results}.

The number of shots in this particular run was 8192, the maximum, and the outcome results are listed in table \ref{tab:bqgb-results}. Clearly the two states we are most interested in, $\ket{001}$ and $\ket{100}$, were the most common outcomes, with 36.51\% and 17.87\% of the outcomes respectively. However neighbouring states $\ket{000}$ and $\ket{010}$ were also very prominent with $> 10\%$ of the overall outcomes each. 

Thus we can claim that our circuit has worked on real hardware, but that the overall noise floor omnipresent in NISQ-era quantum computing poses a real challenge for extended use of multiple peg QGB's constructed in the manner we present. 

\subsection{Locally Simulated Biased-QGB}

\begin{figure}
    \centering
    \includegraphics[width=0.8\textwidth]{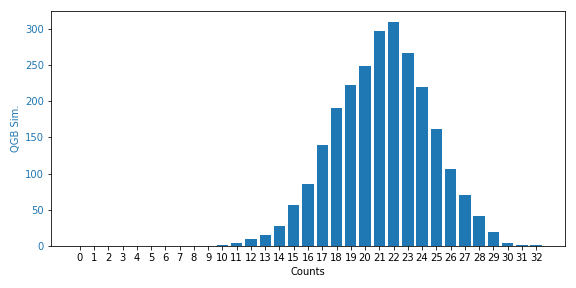}
    \caption{The re-scaled output of the biased 4-QGB locally simulated circuit.}
    \label{fig:bqgb_local_hist}
\end{figure}

We extended our methods in section \ref{sec:qgb-locsim} to run a biased QGB by means of swapping out the $H$ gates for $R_x(\theta)$ gates, and setting $\theta = \frac{2 \pi}{3}$ as described in section \ref{sec:bqpeg-ex}. As before, we present our plotted results in figure \ref{fig:bqgb_local_hist}.

Clearly our scaling produces a desired skewed distribution, which validates our approach in simulations at least. In this run, the mean value produced was 2.66, with a standard deviation of 1.176 and a variance of 1.383.

As before, our OpenQASM is included, found in appendix \ref{appBQGB}.

\section{Fine-Grained Biased QGB Circuits}\label{sec:fine_biased_qgb}

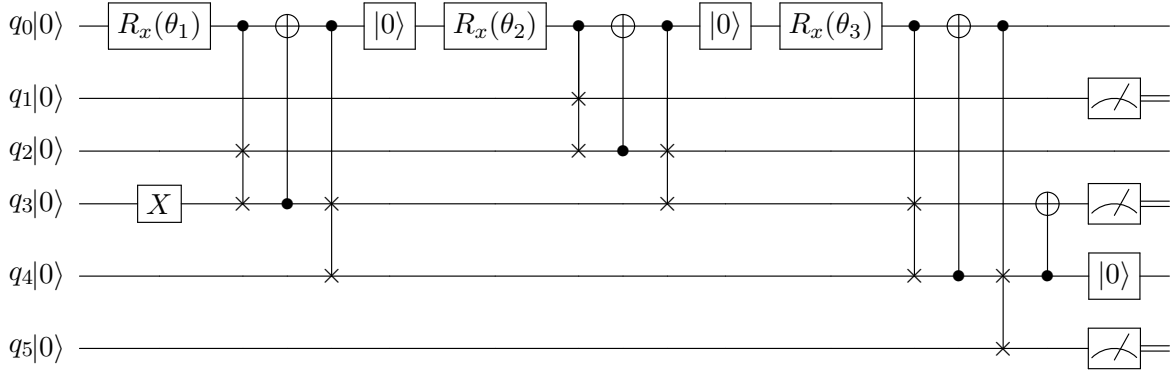
\begin{figure}
    \centering
    \[
    \Qcircuit @C=1em @R=1em {
    & q_0 & 
        & \lstick{\ket{0}} & \gate{R_x(\theta_1)}  & \ctrl{2}      &  \targ     & \ctrl{3}     & \gate{\ket{0}} 
        & \gate{R_x(\theta_2)}  
        & \ctrl{1}    & \targ     & \ctrl{2}    & \gate{\ket{0}} & \gate{R_x(\theta_3)}    & \ctrl{3}      
        & \targ     & \ctrl{4}    & \qw     & \qw  & \qw \\
    & q_1 & 
        & \lstick{\ket{0}} & \qw       & \qw           & \qw        & \qw          & \qw 
        &  \qw      
        & \qswap \qwx & \qw       & \qw         & \qw  & \qw      & \qw           
        & \qw       & \qw    & \qw     & \meter  & \cw \\
    & q_2 & 
        & \lstick{\ket{0}} & \qw       & \qswap        &  \qw       &  \qw         & \qw 
        &  \qw      
        & \qswap \qwx & \ctrl{-2} & \qswap \qwx & \qw  & \qw     & \qw           
        & \qw       & \qw         & \qw     & \qw  & \qw \\
    & q_3 & 
        & \lstick{\ket{0}} & \gate{X}  & \qswap \qwx   &  \ctrl{-3} &  \qswap      & \qw 
        &  \qw      
        & \qw         & \qw       & \qswap \qwx & \qw & \qw & \qswap    
        & \qw       & \qw     & \targ    & \meter  & \cw \\
    & q_4 & 
        & \lstick{\ket{0}} & \qw       & \qw           & \qw        &  \qswap \qwx & \qw 
        & \qw       
        & \qw         & \qw       & \qw         & \qw  & \qw     & \qswap \qwx   
        & \ctrl{-4} & \qswap      & \ctrl{-1}   & \gate{\ket{0}}  & \qw \\
    & q_5 & 
        & \lstick{\ket{0}} & \qw       & \qw           & \qw        & \qw          & \qw
        & \qw      
        & \qw         & \qw       & \qw         & \qw   & \qw    & \qw 
        & \qw       & \qswap \qwx &  \qw & \meter  & \cw 
} \]
    \caption{A Quantum circuit for the fine-grained per-peg control of bias on a QGB. This circuit has the additional $CNOT$ at the end to adjust the outputs. The final reset is not used here but would be important for subsequent rows.}
    \label{fig:quantum_3peg_fine_biased}
\end{figure}

\subsection{Iterating Pegs for Fine-Grained Control on a QGB}

In figure \ref{fig:quantum_3peg_fine_biased} we show how we can have per-peg control of the bias in a QGB. This circuit takes three parameters for the $x$-axis rotations $\theta_1$, $\theta_2$, and $\theta_3$. 

The construction is based on the circuit in figure \ref{fig:quantum_3peg} but with the 10$^{th}$ $CNOT$ gate replaced with a $RESET$ and $R_x$ gate preparing the final peg instead. To adjust for the loss of this correcting $CNOT$ gate we have to use an additional $CNOT$ gate at the end to add the `leftover' rotation to the correct output on $q_3$, followed by an additional $RESET$. Although we do not need this here, it would be required if we were to add additional rows of pegs.

\subsection{Features and Simulation of a Fine-Grained Biased QGB}

\subsubsection{Features}

It is important that the pegs proceed in linear order so that the corrective $CNOT$ gates can function properly. The correction needs to take place for each peg except the first. As such the gate count for an $n$-row fine-grained QGB is roughly:
\begin{itemize}
    \item One $X$ gate for the circuit.
    \item Per peg:
    \begin{itemize}
        \item One $RESET$ gate.
        \item One $R_x(\theta)$ gate.
        \item Two controlled-$SWAP$ gates.
        \item One $CNOT$ gate
    \end{itemize}
    \item For row $i$, $(i-1)$-many $CNOT$ and $RESET$ gates for the end of that row's pegs, for each row.
    \item $n$ measurement gates.
\end{itemize}

Thus overall we have $$ 1 + 5 \Big( \frac{n^2 + n}{2} \Big) + \Big ( \frac{(n-1)^2 + (n-1)}{2} \Big ) + n = 3n^2 + 3n + 1$$

\subsubsection{Simulations}

\begin{figure}
    \centering
    \includegraphics[width=0.6\textwidth,keepaspectratio]{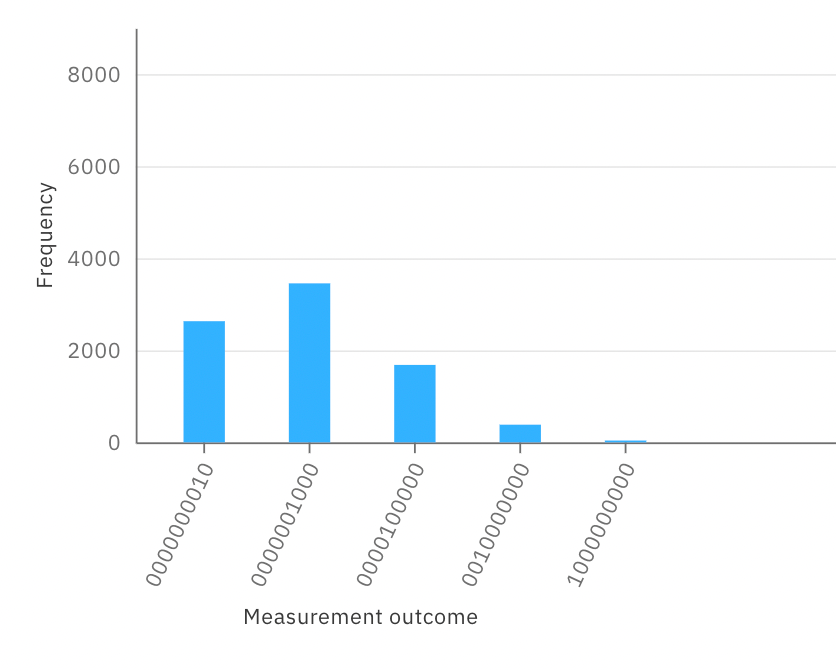}
    \caption{Results histogram for a fine-grained biased QGB simulation. The circuit is depicted in figure \ref{fig:ibm_4QGB}.}
    \label{fig:ibm_fg_biased_results}
\end{figure}

The third (lower) circuit in figure \ref{fig:ibm_4QGB} is our example of a 4-QGB that has control over each peg. Although we could only simulate this circuit, we have also included the OpenQASM in Appendix \ref{appFGBQGB}. 

The results are in figure \ref{fig:ibm_fg_biased_results}. Although we are only simulating a uniform 75\%:25\% across all pegs in the QGB the results match our other simulations. 

This worked example helps to demonstrate the function of the corrective $CNOT$ and $RESET$ gates we mentioned previously. The left-to-right top-to-bottom arrangement of the quantum peg circuit modules means that our corrective additions and resets occurs always on the lower parts of the circuit. 

\section{Conclusions}

We have presented a more intuitive and efficient way of generating quantum Galton boards. We have also shown that these circuits can be simulated effectively, and are reasonably optimal, if of some limited output power. We have also shown how these QGB implementations can be biased to produce other distributions in a reliable way. We have also provided methodologies and code to achieve these results locally, with a view to expanding research on these quantum algorithms.

Clearly noise is a limiting factor for circuits using the Fredkin (controlled-$SWAP$) gates on real hardware owing to the extensive transpiler work that has to take place, significantly increasing the gate count that in turn lifts the noise floor. Until such gates are supported natively or error-corrected qubits are in play, our approach will have limitations.

What we have shown is that the true power of a quantum computer in simulating this kind of behaviour lies in the randomness that allows the outputs to be reliable given the fairly minimal circuit construction in play. Thus, we consider there is value in utilising these QGB circuits as they provide results based on high quality randomness.

Whilst other approaches give $2^n$ possible outputs for $n$ qubit measurements on $\mathcal{O}(n^2)$ resources in play, we note that it is straightforward to transform this normally distributed output to another by linearly increasing numbers of repeated runs of the circuit.

It is a natural fact that, given the current state of NISQ-era quantum computers available to us, the most often thought of route for generating normally-distributed outputs with quantum technology is to get quantum random bits and feed them into a well known sampling algorithm such as the Ziggurat \cite{Marsaglia2000} or others that are in common currency, see \cite[Vol II.]{knuth97}.

Though this route still seems to be the best possible option in practical terms, our aim is that presenting this approach can lend more variety to the construction of quantum circuits that solve the Galton Board simulation and Universal Statistical simulation problems. 

\bibliographystyle{acm}
\bibliography{refs}

\pagebreak
\appendix

\section{OpenQASM for 4-QGB}\label{appA}

We present the OpenQASM code for a 4-level QGB, as simulated on IBM-QX.

\begin{multicols}{2}
\begin{verbatim}
OPENQASM 2.0;
include "qelib1.inc";

qreg q[10];
creg c[10];

reset q[0];
x q[5];
h q[0];
cswap q[0],q[4],q[5];
cx q[5],q[0];
cswap q[0],q[5],q[6];
reset q[0];
h q[0];
cswap q[0],q[3],q[4];
cx q[4],q[0];
cswap q[0],q[4],q[5];
cx q[5],q[0];
cswap q[0],q[6],q[7];
cx q[6],q[0];
cswap q[0],q[5],q[6];
reset q[0];
h q[0];
cswap q[0],q[2],q[3];
cx q[3],q[0];
cswap q[0],q[3],q[4];
cx q[4],q[0];
cswap q[0],q[7],q[8];
cx q[7],q[0];
cswap q[0],q[6],q[7];
cx q[6],q[0];
cswap q[0],q[5],q[6];
cx q[5],q[0];
cswap q[0],q[4],q[5];
reset q[0];
h q[0];
cswap q[0],q[1],q[2];
cx q[2],q[0];
cswap q[0],q[2],q[3];
cx q[3],q[0];
cswap q[0],q[3],q[4];
cx q[4],q[0];
cswap q[0],q[4],q[5];
cx q[5],q[0];
cswap q[0],q[5],q[6];
cx q[6],q[0];
cswap q[0],q[6],q[7];
cx q[7],q[0];
cswap q[0],q[7],q[8];
cx q[8],q[0];
cswap q[0],q[8],q[9];
measure q[1] -> c[1];
measure q[2] -> c[2];
measure q[3] -> c[3];
measure q[4] -> c[4];
measure q[5] -> c[5];
measure q[6] -> c[6];
measure q[7] -> c[7];
measure q[8] -> c[8];
measure q[9] -> c[9];
\end{verbatim}
\end{multicols}

\pagebreak

\section{OpenQASM for a Biased 4-QGB}\label{appBQGB}

We present the OpenQASM code for a 4-level Biased QGB using the circuit change example in section \ref{sec:bqpeg-ex}.

\begin{multicols}{2}
\begin{verbatim}
OPENQASM 2.0;
include "qelib1.inc";

qreg q[10];
creg c[10];

reset q[0];
x q[5];
rx(2*pi/3) q[0];
cswap q[0],q[4],q[5];
cx q[5],q[0];
cswap q[0],q[5],q[6];
reset q[0];
rx(2*pi/3) q[0];
cswap q[0],q[3],q[4];
cx q[4],q[0];
cswap q[0],q[4],q[5];
cx q[5],q[0];
cswap q[0],q[6],q[7];
cx q[6],q[0];
cswap q[0],q[5],q[6];
reset q[0];
rx(2*pi/3) q[0];
cswap q[0],q[2],q[3];
cx q[3],q[0];
cswap q[0],q[3],q[4];
cx q[4],q[0];
cswap q[0],q[7],q[8];
cx q[7],q[0];
cswap q[0],q[6],q[7];
cx q[6],q[0];
cswap q[0],q[5],q[6];
cx q[5],q[0];
cswap q[0],q[4],q[5];
reset q[0];
rx(2*pi/3) q[0];
cswap q[0],q[1],q[2];
cx q[2],q[0];
cswap q[0],q[2],q[3];
cx q[3],q[0];
cswap q[0],q[3],q[4];
cx q[4],q[0];
cswap q[0],q[4],q[5];
cx q[5],q[0];
cswap q[0],q[5],q[6];
cx q[6],q[0];
cswap q[0],q[6],q[7];
cx q[7],q[0];
cswap q[0],q[7],q[8];
cx q[8],q[0];
cswap q[0],q[8],q[9];
measure q[1] -> c[1];
measure q[2] -> c[2];
measure q[3] -> c[3];
measure q[4] -> c[4];
measure q[5] -> c[5];
measure q[6] -> c[6];
measure q[7] -> c[7];
measure q[8] -> c[8];
measure q[9] -> c[9];
\end{verbatim}
\end{multicols}

\pagebreak

\section{OpenQASM for Fine-Grained Biased QGB}\label{appFGBQGB}
\begin{multicols}{2}
\begin{verbatim}
OPENQASM 2.0;
include "qelib1.inc";

qreg q[10];
creg c[10];
reset q[0];
x q[5];
rx(2*pi/3) q[0];
cswap q[0],q[4],q[5];
cx q[5],q[0];
cswap q[0],q[5],q[6];
reset q[0];
rx(2*pi/3) q[0];
cswap q[0],q[3],q[4];
cx q[4],q[0];
cswap q[0],q[4],q[5];
reset q[0];
rx(2*pi/3) q[0];
cswap q[0],q[5],q[6];
cx q[6],q[0];
cswap q[0],q[6],q[7];
barrier q[0],q[1],q[2],q[3],q[4];
barrier q[5],q[6],q[7],q[8],q[9];
reset q[0];
cx q[6],q[5];
rx(2*pi/3) q[0];
reset q[6];
cswap q[0],q[2],q[3];
cx q[3],q[0];
cswap q[0],q[3],q[4];
reset q[0];
rx(2*pi/3) q[0];
cswap q[0],q[4],q[5];
cx q[5],q[0];
cswap q[0],q[5],q[6];
reset q[0];
rx(2*pi/3) q[0];
cswap q[0],q[6],q[7];
cx q[7],q[0];
cswap q[0],q[7],q[8];
barrier q[0],q[1],q[2],q[3],q[4];
barrier q[5],q[6],q[7],q[8],q[9];
reset q[0];
cx q[5],q[4];
cx q[7],q[6];
rx(2*pi/3) q[0];
reset q[5];
reset q[7];
cswap q[0],q[1],q[2];
cx q[2],q[0];
cswap q[0],q[2],q[3];
reset q[0];
rx(2*pi/3) q[0];
cswap q[0],q[3],q[4];
cx q[4],q[0];
cswap q[0],q[4],q[5];
reset q[0];
rx(2*pi/3) q[0];
cswap q[0],q[5],q[6];
cx q[6],q[0];
cswap q[0],q[6],q[7];
reset q[0];
rx(2*pi/3) q[0];
cswap q[0],q[7],q[8];
cx q[8],q[0];
cswap q[0],q[8],q[9];
barrier q[0],q[1],q[2],q[3],q[4];
barrier q[5],q[6],q[7],q[8],q[9];
cx q[4],q[3];
cx q[6],q[5];
cx q[8],q[7];
reset q[4];
reset q[6];
reset q[8];
measure q[1] -> c[1];
measure q[2] -> c[2];
measure q[3] -> c[3];
measure q[4] -> c[4];
measure q[5] -> c[5];
measure q[6] -> c[6];
measure q[7] -> c[7];
measure q[8] -> c[8];
measure q[9] -> c[9];
\end{verbatim}
\end{multicols}

\section{IBM-QX Circuit Diagrams}\label{appB}

To aid understanding, figure \ref{fig:ibm_4QGB} depicts the circuit used for simulation, derived from the OpenQASM code above in Appendices \ref{appA}, \ref{appBQGB}, and \ref{appFGBQGB}.

\begin{sidewaysfigure}
    \centering
    \includegraphics[width=\textwidth,keepaspectratio]{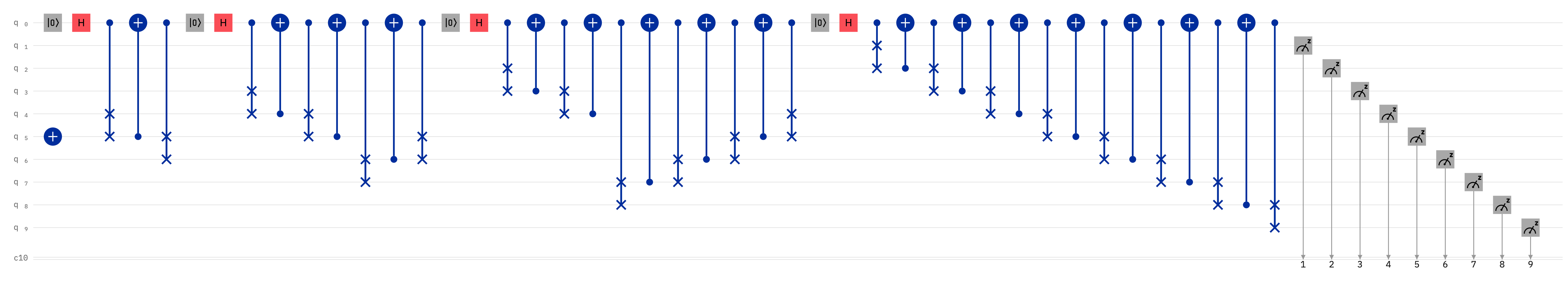}
    \includegraphics[width=\textwidth,keepaspectratio]{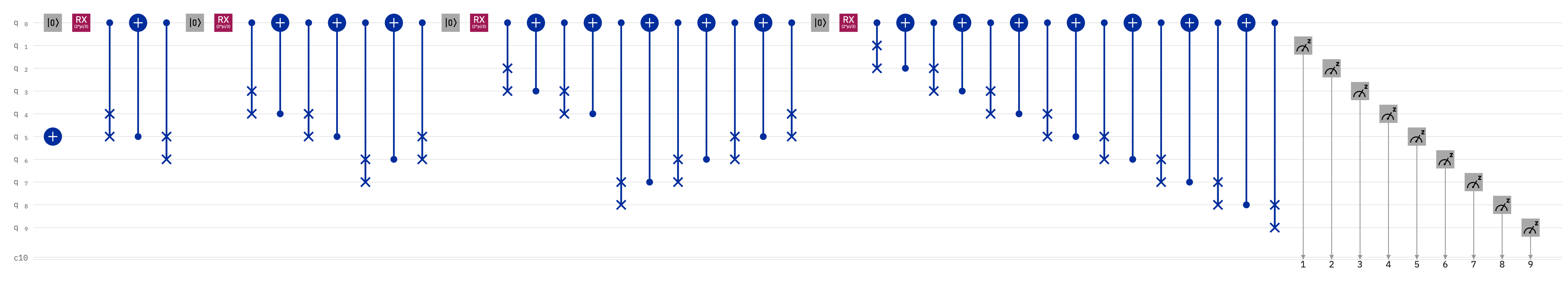}
    \includegraphics[width=\textwidth,keepaspectratio]{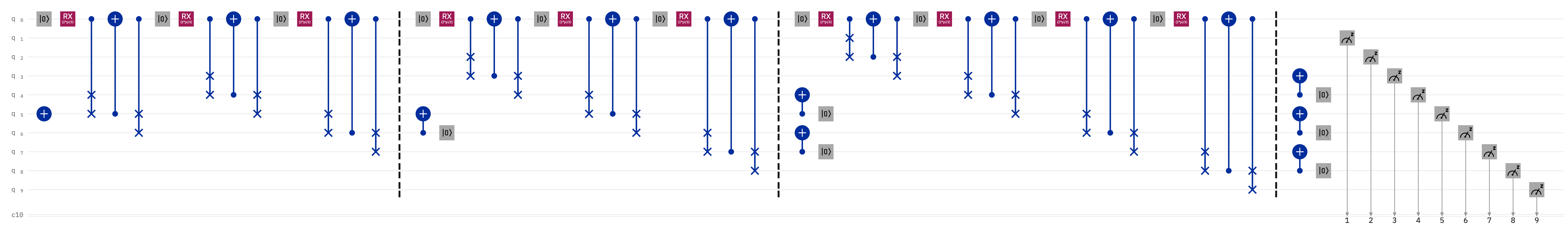}
    \caption{The 4-QGB circuit (upper) and the biased 4-QGB circuit (middle) and fine-grained biased QGB (lower).}
    \label{fig:ibm_4QGB}
\end{sidewaysfigure}

\begin{sidewaysfigure}
    \centering
    \includegraphics[width=\textwidth,keepaspectratio]{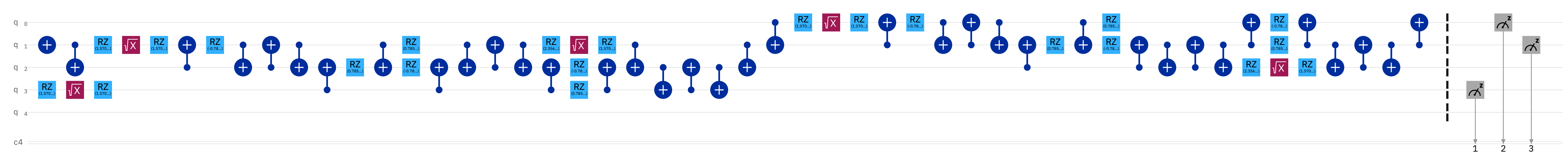}
    \includegraphics[width=\textwidth,keepaspectratio]{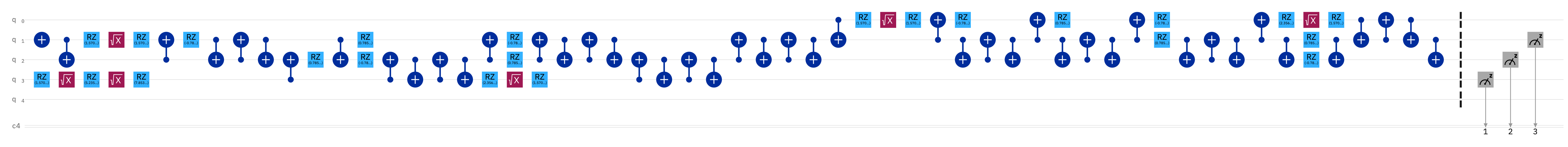}
    \caption{Transpiled Quantum Peg circuits for the ibmq-manila quantum computer. The unbiased (upper) and biased (lower) transpiled circuits are presented.}
    \label{fig:ibm_transpiled_peg}
\end{sidewaysfigure}

\end{document}